\documentstyle[eqsecnum,aps,pre,epsf]{revtex}
\newcommand{\equn}[1]{\begin{equation}\label{#1}}
\newcommand{\eqan}[1]{\begin{eqnarray}\label{#1}}
\newcommand{\bosy}[1]{\bf{#1}}
\newcommand{\eqa}{\begin{eqnarray}}
\newcommand{\equ}{\begin{equation}}
\newcommand{\nuqe}{\end{equation}}
\newcommand{\uqe}{\end{equation}}
\newcommand{\naqe}{\end{eqnarray}}
\newcommand{\aqe}{\end{eqnarray}}
\newcommand{\nonu}{\nonumber}

\newcommand{\goto}{\rightarrow}
\newcommand{\half}{\frac{1}{2}}

\newcommand{\n}{{\bosy \nabla}} 
\newcommand{\e}{{\rm e}}
\newcommand{\Si}{\sigma}
\newcommand{\ka}{\kappa}
\newcommand{\q}{{\bf q}}
\newcommand{\rb}{{\bf r}}
\newcommand{\y}{{\bosy \rho}} 
\newcommand{\dH}{\delta h}
\newcommand{\rw}{{\rm s}}
\newcommand{\as}{\epsilon}
\newcommand{\zs}{z_{\rm s}}
\newcommand{\chip}{(\xi/\xi_\ka)}

\begin{document}

%
%

\title{Supported Membranes on Chemically Structured and Rough Surfaces}
\author{Peter S.\ Swain$^{(1)}$ and David Andelman$^{(2)}$
\\
$^{(1)}$Center for Studies in Physics and Biology, The
Rockefeller University\\
1230 York Avenue, New York, NY 10021\\
$^{(2)}$School of Physics and Astronomy,\\
Raymond and Beverly Sackler Faculty of Exact Sciences,\\
Tel Aviv University, Ramat Aviv 69978, Tel Aviv, Israel}
\maketitle

\begin{abstract}

We present a general linear response description of membrane adhesion at
rough or chemically structured surfaces. Our method accounts for non-local
Van der Waals effects and contains the more approximate (and local)
Deryagin approach in a simple limit. Specializing to supported membranes
we consider the effects of substrate structure on the membrane adhesion
energy and configuration. Adhesion is usually less favorable for rough
substrates and the membrane shape tends to follow that of the surface
contours. Chemical patterning, however, favors adhesion with the membrane
configuration being out of phase with the surface structure. Finally,
considering a surface indented with `V'-shaped trenches, we show that our
approach is in fair agreement with an exact numerical solution.

\end{abstract}

\section{Introduction}

Supported membranes strongly adhere to substrates and lie typically at
separation distances of between $10\, {\rm \AA}$ and $40 \, {\rm \AA }$
\cite{sack}. Such small values have lead to their adoption by the
biotechnology industry \cite{sack2} and, in particular, given them an
important role in the development of biosensors. Supported membranes
enable one to biofunctionalize an inorganic surface and can provide an
ultra-thin, highly electrically resistant layer on top of a conducting
substrate. They provide a means of immobilizing proteins with a
well-defined orientation and in a non-denaturing environment \cite{sala}.
If these proteins are receptors then one can use electrical or optical
means to detect or ``sense'' the binding of ligands to the receptors
\cite{jenkins}.

Supported membranes can be formed by the spreading of a bilayer over a
substrate, vesicle fusion taking place at a substrate or by lipid
monolayer transfer using a Langmuir-Blodgett technique \cite{sack}.
However, in nearly all applications the substrates used are not simply
planar and homogeneous but are patterned, either chemically
\cite{jenkins,groves} or geometrically \cite{osborn,ogier}. The theory of
membrane adhesion has typically concentrated on adhesion at ideal planar
surfaces \cite{seiflip}. In a recent paper \cite{us}, we have provided a
simple description of the adhesive properties of membranes at rough
surfaces. In this article, we would like to present a more general
approach to membrane adhesion which includes the possibility of chemical
patterning.

We begin with a summary of the basic assumptions of our model and of the
intermolecular interactions involved. In Sec.\ \ref{planarand}, a planar,
homogeneous substrate is considered as the starting point for the linear
response theory that follows. We show how the simpler approach of Ref.\
\cite{us} is included in the present work and then proceed to consider
several illustrative examples in Sec.\ \ref{results1}. Finally in Sec.\
\ref{results2}, our analytical description is compared and contrasted with
a complete numerical solution. Throughout, we emphasize the effect of
substrate roughness and chemical heterogeneities on the adhesive
properties of the supported membrane.

\section{The free energy}

To begin we consider a membrane supported at a substrate that can be
either geometrically structured (non-planar), Fig.\ \ref{fig1}(a), or
chemically structured (patterned with different chemical compounds), Fig.\
\ref{fig1}(b) and (c). For example, a surface can be chemically structured
by depositing different chemical layers, see Fig.\ \ref{fig1}(b), or
adjoining different chemical surfaces together to make a columnar
structure, Fig.\ \ref{fig1}(c). For a membrane adhering to such a surface
this patterning will greatly influence the membrane configuration and
adhesion energy. Inspired by advances in the theory of wetting
\cite{david}, we adopt a general mean-field
 approach in which the configuration taken up by the membrane is
 one that minimizes the free energy. In order to find this optimum
 configuration we first discuss the form of the free energy functional.

If the membrane has an elastic modulus $\ka$ and tension $\Si$,
 then its bending energy can be described by the functional \cite{canham}
\equ
\int dS \, \sqrt{g} \, \left[ \sigma + \half \kappa (2H)^2
\right]
\uqe
where the integral is over the membrane surface, $g$ is the determinant of
the metric, $H=(c_1+c_2)/2$ the mean curvature, and $c_1$ and $c_2$ the
two principal curvatures. We have ignored here the Gaussian curvature
contribution as only a membrane with a fixed topology (flat on large
lengthscales and of infinite size) is considered \cite{gb}. Throughout the
paper we choose to work in the Monge representation. Letting $\y = (x,y)$
be a two dimensional planar vector, the heights of the surface and
membrane above some reference $\y$-plane are $z_\rw(\y)$ and $h(\y)$,
respectively (see Fig.\ \ref{fig1}).

To account for the interaction of the membrane with the substrate, we
include a potential term, $V(h; \zs, \as)$, in the free energy. As
already mentioned, $\zs(\y)$ accounts for the substrate's geometrical
structure and describes its surface configuration, while
$\as({\bf r})$ denotes any chemical inhomogeneities. The potential
can have a number of different components \cite{lip,isr}. For our
case, the most important of these is the Van der Waals contribution,
which is given by
\equn{pot0}
V_{\rm vdw}(h;\zs,\as) = - \Bigl[ W(h;z_\rw,\as) - W(h+\delta;z_\rw,\as) \Bigr]
\nuqe
where $\delta$ is the membrane thickness and is typically around $40$ \AA.
Due to the bilayer nature of the membrane, two terms involving the Van der
Waals potential, $W(h;z_\rw,\as)$, are necessary; in particular for
supported membranes where $\delta \approx h$.

For a thin fluid film of thickness $h(\y)$ resting on an inhomogeneous
solid, one can sum over all possible pair interactions between the
molecules in the upper half space, capped from below by the surface
$z=h(\y)$, and those in the lower half space, capped from above by
$z=z_\rw(\y)$, to show that $W(h;z_\rw)$ satisfies \cite{david}
\equn{pot}
W(h;z_\rw,\as) = \int_{h(\y)}^\infty dz \int d^2 \y'
\int_{-\infty}^{z_\rw(\y')} dz'\,\, w_0({\bf r}-{\bf r}') \Bigl[ 1 +
\as({\bf r}') \Bigr]
\nuqe
with
\equ w_0({\bf r}) = \frac{{A_0}}{\pi^2} \left( \frac{1}{r^6}
\right) \label{vdw} \uqe

The latter models non-retarded Van der Waals interactions.
Equation (\ref{pot}) contains a position dependent
Hamaker constant
\equn{Afull} A({\bf r}) = {A_0} \Bigl[ 1 + \as({\bf r}) \Bigr]
\nuqe
with ${A_0}$ the average value
\equ {A_0} = \frac{\int d^2 \rb \, A(\rb)}{\int d^2 \rb} \uqe
and $\epsilon$ the (small) deviation around this average,
\equ \as(\rb) = \frac{A(\rb)-A_0}{A_0} \uqe
We emphasize that $W(h;\zs,\as)$ is a functional of both $\zs(\y)$
and $\as({\bf r})$. 

If $h(\y)$ is set to a constant value, say $h_0$, and both $\zs$
and $\as$ vanish, then (\ref{pot}) becomes the familiar
\equn{vdwflat} W(h_0;0,0) \equiv W_0(h_0) = \frac{A_0}{12\pi}
\cdot \frac{1}{h_0^2} \nuqe
(see Ref.\ \cite{isr}), which is just the Van der Waals potential between
two planar semi-infinite bodies held a distance $h_0$ apart.

Equation (\ref{pot}) provides an attractive interaction and, for
the case of a supported membrane, this is chiefly balanced by
hydration forces. The hydration potential obeys
\equn{hyd}
V_{\rm hyd}(h;z_\rw) = b \, \e^{-\alpha (h-z_\rw)}
\nuqe
where $b$ has units of surface tension and $\alpha$ is an inverse length
of typical size $\alpha^{-1} \simeq 2$--$3 \, $\AA. Due to the very short
range nature of the hydration interaction, we include the dependence on
the substrate structure with a simple local approximation and so $V_{\rm
hyd}$ is just a function of the local height $h(\y)-\zs(\y)$. The origin
of hydration forces is still under debate \cite{isr} but they are
generally believed to have some steric contribution. Consequently, while
$b$ in general is position dependent we believe that this is a relatively
minor effect and so choose to keep the simple form of (\ref{hyd}).

The total potential is then
\equn{nochein}
V(h; z_\rw, \as) = V_{\rm vdw}(h; z_\rw, \as) + V_{\rm hyd}(h; z_\rw)
\nuqe
though one could consider more complicated scenarios
involving, for example, electrostatic forces.
Summing all these contributions we can write the total free energy as
\equn{f} {\cal F}[h] = \int d^2 \y \Biggl \{ \sqrt{1+(\n h)^2}
\left[ \Si + \frac{\ka}{2} \left( \vec{{\bosy \n}} \cdot
\frac{\vec{{\bosy \n}} h}{\sqrt{1+(\n h)^2}} \right)^2 \right] +
V(h;z_\rw,\as) \Biggr \} \nuqe
where we have explicitly written out the curvature and tension terms in
the Monge representation.

One of the most relevant quantities in experiments is the membrane
adhesion energy. Within our general mean-field approach, the optimal
height of the membrane is that which minimizes (\ref{f}) and the
value of the free energy when the membrane takes up this
optimum configuration, ${\cal F}_{\rm min}$, leads to a natural
definition of the adhesion energy per unit area
\equn{ad} U \equiv - \left( \frac{{\cal F}_{\rm min}}{S_0} - \Si
\right) \nuqe
Here, $S_0$ is the total area of the projected reference $\y$-plane, $S_0
= \int d^2 \y$ and we have subtracted off a membrane tension term. Doing
so conveniently defines the adhesion energy so that a completely flat
membrane does not have a tension dependent contribution; for a membrane
infinitely far from the surface $U$ will then vanish. Notice that
(\ref{ad}) implies that an attractive surface will have a positive
adhesion energy.

\section{Planar and Homogeneous Substrates}
\label{planarand}

Our results are obtained by analytically expanding the free energy
around its value taken for a planar, chemically homogeneous
substrate. Therefore, we briefly review the results for such an ideal
surface.

For this case, $\epsilon=z_\rw=0$, and the Van der Waals
interaction (\ref{pot}) simplifies to (\ref{vdwflat}), and so
\equn{vvdw}
V_{\rm vdw}(h;0,0) = -\Bigl[ W_0(h) - W_0(h+\delta) \Bigr]
\nuqe
where throughout we use the subscript zero to denote adhesion
at both chemically homogeneous and flat substrates.
Here, $a$ is a fundamental lengthscale in our problem
\equ a = \left( \frac{A_0}{2\pi \Si} \right)^\half \uqe
and is provided by the ratio of the Hamaker constant,
see (\ref{vdwflat}), and the membrane tension \cite{pgg}.

We find that the membrane adopts a flat
configuration, $h(\y) = h_0$, which obeys
\equn{pla1}
\frac{\partial V}{\partial h} = V_{\rm vdw}'(h_0) - \alpha b \,
\e^{-\alpha h_0} = 0
\nuqe
from (\ref{hyd}).

The adhesion energy in this case is simply given as the negative
of the interaction potential. From (\ref{f}), ${\cal F}_{\rm
min} = S_0 \left[ \Si + V(h_0;0,0) \right]$ and so (\ref{ad})
implies that
\eqan{pla2}
U_0 &=& -V(h_0;0,0) \nonu \\
&=& -\left[ V_{\rm vdw}(h_0) + b \, \e^{-\alpha h_0} \right]
\naqe
By definition, $U_0$ is positive for all sufficiently
attractive potentials, $V$. Equations (\ref{pla1})
and (\ref{pla2}) provide the fundamental quantities
upon which our perturbation theory will be built.

In order to allow (semi-quantitative) comparison with experiment and to
give some idea of the magnitude of the quantities involved, we would now
like to specialize to a particular choice of our model parameters (we opt
again for those chosen in Ref.\ \cite{us}), see Table \ref{tab}.  Typical
experimental values of $\Si$ and $\kappa$ are $1.7 \times 10^{-5 } \; {\rm
Jm^{-2}}$ and $35 T$, respectively \cite{rad}. We set the Boltzmann
constant to unity and so at room temperature $T = 4.1 \times 10^{-21} \;
{\rm J}$. Choosing, $A_0 = 2.6 \times 10^{-21} \; {\rm Jm^{-2}}
 \simeq 0.63\,T$ \cite{rad},  implies that the
 lengthscale $a \simeq 49.3 \;$\AA \hspace*{0.5mm}
 and, from (\ref{pla1}), $h_0 \simeq 0.61a\simeq 30
 {\rm \;}$\AA \hspace*{0.5mm} in agreement with
  measured values using specular reflection of neutrons \cite{shirl}. 
The two parameters used here to specify the hydration force, see
(\ref{hyd}), are
\equn{correct}
\begin{array}{lcr}
b = 0.93 \; {\rm Jm^{-2}} &;& \alpha^{-1} = 2.2 \; \mbox{\AA}
\end{array}
\nuqe
which are in accordance with those measured in Ref.\ \cite{pars}.

The potential experienced by the membrane, (\ref{nochein}), is sketched in
Fig.\ \ref{fig2}. From (\ref{pla2}), one can see that
\equn{Ustr}
U_0 \simeq 0.298 \Si = 5.07 \times 10^{-6} \; {\rm Jm^{-2}}
\nuqe

At this point, it is also worth discussing the other lengthscales
which will appear in our treatment.  Defining $v$ as the second
derivative of the potential calculated at the minimum, $h=h_0$,
\eqan{barv}
v &=& \left. \frac{\partial^2}{\partial h^2} V(h;0,0)
\right|_{h=h_0} \nonu \\
&=& V''_{\rm vdw}(h_0;0,0) + V''_{\rm hyd}(h_0)
\naqe
several correlation lengths can be extracted
\equn{corrs}
\begin{array}{lcl}
\xi_\Si^2 = \Si/v &;& \xi_\ka^4 = \kappa /v
\end{array}
\nuqe
together with
\equn{xidefn} \xi^2 = {\ka}/{\Si} = \xi^4_\ka \xi^{-2}_\Si \nuqe
which describes the crossover between the tension and the
rigidity dominated regimes. Their values for the experimental
scenario described above are given in Table \ref{tab}
(which also lists all the other model parameters).

\section{Linear Response Theory}

To carry out a perturbation theory for rough and heterogeneous substrates,
we assume that $\zs$ and $h-h_0$ are small, i.e. $\zs \sim h-h_0 \ll h_0$,
and that any of their products and derivatives are also small. A
(functional) Taylor expansion is then performed which is a fairly
standard, if long, calculation.

To simplify our presentation and ease the algebra, we will assume that the
chemical structure is such that $\as$ can be factorized, i.e.\
\equn{asfac}
\as({\bf r}) = \phi(\y) \, g(z)
\nuqe
for some functions $\phi$ and $g$ and the substrate surface is given by
$z_{\rm s}=0$. Such a factorization while including the layered (constant
$\phi$) and columnar (constant $g$) structures shown in Fig.\ \ref{fig1}
does prevent us from considering surfaces which are both rough and
chemically inhomogeneous. Consequently from this point on, we will
specialize to either rough or chemically patterned substrates.

A few more definitions are in order; first of all, we notice that $v$,
given by (\ref{barv}), can also be shown to obey
\equ
v = V''_{\rm hyd}(h_0) - \int d^2 \y' \Bigl\{ w_0(\y',h_0) -
w_0(\y',h_0+\delta) \Bigr\}
\uqe
using (\ref{pot0}) and (\ref{pot}). The kernel functions
(this choice of nomenclature will become clear later)
\equn{K}
K(\y) = -\frac{1}{v} \left[ w_0(\y,h_0)-w_0(\y,h_0+\delta) \right]
\nuqe
and
\equn{G}
G(\y) = -\frac{1}{v} \int_{h_0}^{h_0+\delta} dz \, g(h_0-z) w_0(\y,z)
\nuqe
will also prove useful.

Then, expanding ${\cal F}[h]$ in (\ref{f}) to second order and
taking advantage of (\ref{pla1}), we find
\eqan{f2} 
\frac{1}{\Si}{\cal F} &\approx& \frac{1}{\Si} {\cal
F}_0 + \frac{1}{2\xi_\Si^2} \int d^2 \y \Biggl\{ \xi_\Si^2 (\n h)^2 +
 \xi_\ka^4 (\n^2 h)^2 +(h-h_0)^2 \nonu \\
& & - 2 (h-h_0)
\Bigl[ {\cal V} \zs + \int d^2 \y' \, K(\y') \zs(\y+\y') \Bigr] +
\zs^2(\y) \Biggr\}
\naqe
for rough surfaces and
\eqan{f2c}
\frac{1}{\Si}{\cal F} &\approx& \frac{1}{\Si} {\cal
F}_0 + \frac{1}{2\xi_\Si^2} \int d^2 \y \Biggl\{ \xi_\Si^2 (\n h)^2 +
 \xi_\ka^4 (\n^2 h)^2 +(h-h_0)^2 \nonu \\
& & - 2 (h-h_0) \Bigl[ \int d^2 \y'\, G(\y')\phi(\y+\y') \Bigr] \Biggr\}
\naqe
for chemical structure, where ${\cal F}_0 \equiv {\cal F}[h_0]$ is the
$h$-independent term in the expansion,
\equn{f0'shere} 
{\cal F}_0 =  \Si \left[ 1 - \frac{1}{\Si}{U_0} \right] S_0
\nuqe
In the case of chemical patterning, ${\cal F}_0$ contains an additional term
\equ
 \int_{-\infty}^0 dz \, V'_{\rm vdw}(h_0-z) \int d^2 \y \, \as(\y,z)
\uqe
which can be made to vanish by choosing the $\y$-plane such that
\equ
\int d^2\y \, \phi(\y) = 0
\uqe
For rough substrates the $\y$-plane is chosen so that $\langle \zs \rangle
= \int d^2\y \, \zs=0$. Here, ${\cal V}$ is given by
\equn{calv}
{\cal V} = \frac{V''_{\rm hyd}(h_0)}{v}
\nuqe

To find the optimum profile, we need to minimize (\ref{f2}) or (\ref{f2c})
 with respect to $h(\y)$, i.e.\ $\delta {\cal F}/ \delta h=0$. 
The resulting Euler-Lagrange equation
 is non-local in $\y$ but linear in $h$,
\equn{el}
\Bigl[ \xi_\ka^4 \n^4  - \xi_\Si^2 \n^2  + 1 \Bigr] \Bigl( h(\y) -
h_0 \Bigr)= \left\{ \begin{array}{ll} \int d^2 \y' \Bigl[ K(\y'-\y) \zs(\y')
 \Bigr] + {\cal V} \zs(\y) & \mbox{\hspace*{8mm} rough} \\
\int d^2 \y' \Bigl[ G(\y'-\y) \phi(\y') \Bigr] & \mbox{\hspace*{8mm}
chemical} \end{array} \right.
\nuqe
where the role that $K$ and $G$ play as kernel functions becomes clear.
Equation (\ref{el}) is the starting point for our linear response
profiles. It is a fourth order non-homogeneous linear differential
equation where the heterogeneity of the substrate enters in the
non-homogeneous term. Due to its linear nature, the solution can be
written down in Fourier space. Defining for any function $f(\y)$ its
Fourier transform
\equ
\tilde{f}(\q) = \int d^2 \y f(\y) \e^{-i \q.\y}
\uqe
we find, via the convolution theorem, that $\delta h(\y) = h(\y)-h_0$ obeys
\equn{elsoln}
\tilde{\delta h}(\q) = \left\{ \begin{array}{ll} \frac{\Bigl[
\tilde{K}(\q) + {\cal V} \Bigr] \tilde{\zs}(\q)}{1+\xi_\Si^2 q^2 +
\xi_\ka^4 q^4} & \mbox{\hspace*{8mm} rough} \\
 & \\
\frac{\Bigl[ \tilde{G}(\q) \tilde{\phi}(\q)\Bigr] }{1+\xi_\Si^2 q^2 +
\xi_\ka^4 q^4} & \mbox{\hspace*{8mm} chemical}
\end{array} \right.
\nuqe

The Fourier transform $\tilde{K}(\q)$ of the kernel function $K(\y)$
 can be calculated using the result
\equ
\int d^2 \y \, \frac{\e^{-i \q.\y}}{(\y^2+h^2)^{m+1}} = 2\pi \left(
\frac{q}{2h} \right)^m \cdot \frac{K_m(q h)}{\Gamma(m+1)}
\uqe
where $K_m(x)$ is the modified Bessel function of the second kind
of order $m$. Non-retarded Van der Waals interactions are
obtained by setting $m=2$ in the above equation (from
(\ref{vdw})). Then the kernel becomes
\equn{kernel2}
\tilde{K}(\q)=\tilde{K}(q) = -\frac{\xi_\Si^2 q^2 a^2}{2} \left[
\frac{K_2(q h_0)}{h^2_0} - \frac{K_2(qh_0 +q\delta)}{(h_0+\delta)^2}
\right]
\nuqe
Notice that in the limit of $q$ tending to zero,
$\lim_{q \goto 0} q^2 K_2(q) = 2$, which implies that
\eqan{Klimit}
\tilde{K}(0) &=& - \xi_\Si^2 a^2 \left( \frac{1}{h^4_0} -
\frac{1}{(h_0+\delta)^4} \right) \nonu \\
&=& 1-{\cal V}
\naqe
from (\ref{barv}) and (\ref{calv}).

Similarly, $\tilde{G}(\q)$ satisfies
\equn{tG}
\tilde{G}(\q) = -\frac{\xi_\Si^2 a^2 q^2}{2} \int_{h_0}^{h_0+\delta}
dz \, \frac{g(h_0-z)}{z^2} K_2(q z)
\nuqe
which then implies that
\equn{Glimit}
\tilde{G}(0) = -\xi_\Si^2 a^2 \int_{h_0}^{h_0+\delta}dz \, \frac{g(h_0-z)}{z^4}
\nuqe
For the case when $g(z) \equiv 1$ (a columnar solid)
\equn{G01}
\tilde{G}(0) = -\frac{\xi_\Si^2 a^2}{3} \left(\frac{1}{h_0^3}
-\frac{1}{(h_0+\delta)^3}\right) < 0
\nuqe
which will be needed in Sec.\ \ref{results1}.

We can also calculate the adhesion energy in Fourier space by Fourier
transforming (\ref{f2}). Using the solution of the Euler-Lagrange
equation, (\ref{elsoln}), and the definition (\ref{ad}) then gives
\equn{ad2}
\frac{U}{\Si} = \left\{ \begin{array}{ll} \frac{U_0}{\Si} - \frac{1}{2S_0}
\int \frac{d^2 \q}{(2\pi \xi_\Si)^2} \left( |\tilde{\zs}(\q)|^2 -
\frac{\left| \Bigl(\tilde{K}(\q)+ {\cal V} \Bigr) \tilde{\zs}(\q)
\right|^2}{1+q^2 \xi_\Si^2 + q^4 \xi_\ka^4} \right) & \mbox{\hspace*{8mm}
rough} \\
 & \\
\frac{U_0}{\Si} + \frac{1}{2S_0} \int \frac{d^2 \q}{(2\pi \xi_\Si)^2} \left(
\frac{\left| \tilde{G}(\q) \tilde{\phi}(\q)
\right|^2}{1+q^2 \xi_\Si^2 + q^4 \xi_\ka^4} \right) & \mbox{\hspace*{8mm}
chemical}
\end{array} \right.
\nuqe

\section{The Deryagin approximation}

In a previous paper \cite{us}, we looked quite extensively at the
adsorption of membranes on
rough substrates and throughout used a
Deryagin-like local approximation \cite{dery1}. In this section we would like
to show how this is included in our present, more general, linear
response approach.

The Van der Waals potential, (\ref{pot0}), due to its functional
(non-local) dependence on the inhomogeneities, provides most of the
difficulties in any analytical method. These complications are neatly
circumvented by the Deryagin approximation. Equation (\ref{pot}) can, by a
simple change of variable, be re-written as
\equn{pot1}
W(h;z_\rw,\as) = \left\{ \begin{array}{ll} \int_{h(\y)}^\infty dz \int d^2
\y' \int_{-\infty}^{z_\rw(\y+\y')} dz'\, w_0(\y',z-z') &
\mbox{\hspace*{8mm} rough} \\

\int_{h(\y)}^\infty dz \int d^2 \y' \int_{-\infty}^0 dz'\, w_0(\y',z-z')
\Bigl[ 1+\phi(\y+\y') g(z') \Bigr] & \mbox{\hspace*{8mm} chemical}
\end{array} \right.
\nuqe
where we have adopted (\ref{asfac}). The Deryagin method amounts
to replacing
\equn{dodery}
\begin{array}{ll}
\zs(\y+\y') \goto \zs(\y) & \mbox{\hspace*{8mm} rough} \\
\phi(\y + \y') \goto \phi(\y) & \mbox{\hspace*{8mm} chemical} \\
\end{array}
\nuqe
which removes the functional character of (\ref{pot1}) and so
neglects almost all non-local effects (we still retain the
integral over $g(z')$). Once (\ref{dodery}) has been performed,
one can (Taylor) expand the free energy in powers of $h-h_0$
and $\zs$ as before.

However, such an approach turns out to be equivalent to replacing the
kernel functions, (\ref{K}) and (\ref{G}), in the linear response theory
by Dirac delta functions,
\eqan{spikes}
K(\y - \y') &\goto& \tilde{K}(0) \delta(\y-\y') \nonu \\
G(\y - \y') &\goto& \tilde{G}(0) \delta(\y-\y')
\aqe
which transparently shows the local character of the Deryagin
technique. Equation (\ref{spikes}) implies that the Fourier
transforms of the kernel functions are now simply constants.

Consequently, the Euler-Lagrange equation, (\ref{el}), becomes
\equn{el2}
(\xi_\kappa^4 \n^4-\xi_\Si^2 \n^2 + 1) \delta h(\y) = \left\{
\begin{array}{ll} \zs(\y) & \mbox{\hspace*{8mm} rough} \\ \tilde{G}(0)
\phi(\y) & \mbox{\hspace*{8mm} chemical} \\
\end{array} \right.
\nuqe
where we have used (\ref{Klimit}). In Fourier space the solution is
\equn{deryel}
\tilde{\delta h}(\q) = \left\{ \begin{array}{ll} \frac{ \tilde{\zs}(\q)
}{1+\xi_\Si^2 q^2 + \xi_\ka^4 q^4} & \mbox{\hspace*{8mm} rough} \\
 & \\
\frac{\tilde{G}(0)
\tilde{\phi}(\q)}{1+\xi_\Si^2 q^2 + \xi_\ka^4 q^4} & \mbox{\hspace*{8mm}
chemical} \end{array} \right.
\nuqe
with an adhesion energy given by (\ref{ad2}) with the substitution
(\ref{spikes}),
\equn{deryad} 
\frac{U}{\Si} = \left\{ \begin{array}{ll} \frac{U_0}{\Si} - \frac{1}{2S_0}
\int \frac{d^2 \q}{(2\pi \xi_\Si)^2} \Biggl(
 \frac{ (q^2 \xi_\Si^2 + q^4 \xi_\ka^4) \left| \tilde{\zs}(\q) \right|^2}
 {1+q^2 \xi_\Si^2 + q^4 \xi_\ka^4} \Biggr) & \mbox{\hspace*{8mm} rough} \\
 & \\
\frac{U_0}{\Si} + \frac{1}{2S_0}
\int \frac{d^2 \q}{(2\pi \xi_\Si)^2} \Biggl( 
 \frac{\left| \tilde{G}(0) \tilde{\phi}(\q) \right|^2}
 {1+q^2 \xi_\Si^2 + q^4 \xi_\ka^4} \Biggr)  & \mbox{\hspace*{8mm} chemical}
\end{array} \right.
\nuqe
For the case of roughness, these are exactly the results obtained in our
previous paper \cite{us}. We note that there is a difference in sign
between the two cases; while $\delta U \equiv U-U_0 <0$ for the rough
case, $\delta U > 0$ for chemically patterned surfaces. This is one of our
main observations and will be discussed in Sec.\ \ref{comparisonchem}.

\section{Predictions of the Linear Response Method}
\label{results1}

The non-local perturbation method is embodied by (\ref{el}) and
(\ref{ad2}). In this section we compare and contrast this approach with
the simpler local Deryagin approximation. As both methods are based on
perturbation theory, we have the caveat that
\equn{perturbvalid}
U/U_0 \approx 1
\nuqe
for the method to be valid, i.e.\ that the perturbation correction
is much smaller than the term it improves on. The fundamental
difference between the approaches is in the treatment of the Van
der Waals potential; in the Deryagin approximation there is only a purely local
attraction while the linear response includes non-local collective
effects. The Euler-Lagrange equations, (\ref{el}) and (\ref{el2}),
most clearly illustrate this.

\subsection{Chemically structured surfaces}

For the types of structure considered, (\ref{f2}) gives the general linear
response free energy. One can define a bending (BE) and potential energy
(PE) contribution to this by simply letting the bending energy be that
part which vanishes when $\sigma$ and $\kappa$ are set to zero and the
potential energy that part that remains. Then, using the solution,
(\ref{elsoln}), we find
\equn{chemPE}
{\rm PE}_{\rm chem}=  -\int \frac{d^2 \q}{(2\pi)^2} \,
\frac{1+2Q}{(1+Q)^2} |\tilde{G}(\q) \tilde{\phi}(\q)|^2
\nuqe
and
\equn{chemBE}
{\rm BE}_{\rm chem} =  \int \frac{d^2 \q}{(2\pi)^2} \,
\frac{Q}{(1+Q)^2} | \tilde{G}(\q) \tilde{\phi}(\q)|^2
\nuqe
with $Q= \xi_\Si^2 q^2 + \xi_\ka^4 q^4$ (a constant contribution,
$V(h_0;0,0)$,
 has been ignored in the potential energy term).  Here and throughout the
rest of the paper, we fix $g(z) \equiv 1$ for clarity and so consider only
columnar solids. The reader interested in layered solids should consult
Ref.\ \cite{david} whose results can be simply extended to membranes.

Using the fact that
\equn{Glimits}
0 > \tilde{G}(q) \ge \tilde{G}(0)
\nuqe
for all $q$, and remembering that the Deryagin approximation is recovered
when the kernel function, $\tilde{G}(q)$, is set to its value at $q=0$,
one can see almost by inspection that the membrane potential energy will
be higher than in the Deryagin case while the bending energy will be
lower. In fact, the increment in the potential energy will be greater than
the bending energy decrement and we can therefore expect the surface to
appear less attractive, due to non-local effects, with a correspondingly
lower $U$.

A sinusoidally patterned surface, translationally invariant in the
$y$-direction, is the simplest choice with which to illustrate this
behavior. Setting
\equ
\phi(\y) = \Lambda_{\rm c} \, \sin(p_{\rm c} x)
\uqe
with $\Lambda_{\rm c}$ being the amplitude of the Hamaker coefficient
oscillations and $2\pi/p_{\rm c}$ their period, (\ref{elsoln}) then
implies
\equ
 h(\y) = h_0 + \frac{ \Lambda_{\rm c} \tilde{G}(p_{\rm
c})
 \sin(p_{\rm c} x)}{1+\xi_\Si^2 p_{\rm c}^2 + \xi_\ka^4 p_{\rm c}^4}
\uqe
where $\tilde{G}(p_{\rm c})<0$ from (\ref{Glimits}). The adhesion energy
is also easily calculated
\eqan{adsiny}
\frac{U}{\Si}= \frac{U_0}{\Si} + \frac{1}{4\xi_\Si^2} \cdot
\frac{\Lambda_{\rm c}^2 |
 \tilde{G}(p_{\rm c})|^2}{1+p_{\rm c}^2\xi_\Si^2 + p_{\rm c}^4 \xi_\ka^4}
\naqe

In Fig.\ \ref{fig3}(a), this adhesion energy is plotted against wavenumber
$p_{\rm c}$. One can see that there is good agreement between the Deryagin
and linear response approaches. As the period of the sinusoidal variation
increases the membrane is less and less able to respond to the structural
variations and for high $p_{\rm c}$ the adhesion energy takes its planar
value once more. Any chemical structure has been effectively washed away.

\subsection{Rough surfaces}

In this case, the two contributions to the adhesion energy are
\equn{PErough}
{\rm PE}_{\rm rough} = \int \frac{d^2 \q}{(2\pi)^2} \, |\tilde{\zs}(\q)|^2
\left\{ 1 - |\tilde{K}(\q)+{\cal V}|^2 \frac{1+2Q}{(1+Q)^2} \right\}
\nuqe
and
\equn{BErough}
{\rm BE}_{\rm rough} =  \int \frac{d^2 \q}{(2\pi)^2} \, \frac{Q}{(1+Q)^2} |
(\tilde{K}(\q) + {\cal V}) \tilde{\zs}(\q)|^2
\nuqe
As
\equn{Klimits}
\tilde{K}(q)+{\cal V} \ge \tilde{K}(0)+{\cal V} = 1
\nuqe
for all $q$, the complete opposite behavior results with the
bending energy increased by the non-local contributions and the
potential energy decreased below the decrement to the bending
energy. The surface becomes more attractive and the adhesion energy
inceases above the Deryagin result.

Looking at a chemically homogeneous but geometrically corrugated surface
we choose
\equ
\zs(\y) = \Lambda_{\rm s} \, \sin(p_{\rm s} x) \\
\uqe
so that the surface corrugations have an amplitude of $\Lambda_{\rm s}$
and a period of $2\pi/p_{\rm s}$. Equation (\ref{elsoln}) gives
\equ
h(\y) = h_0 + \frac{\Lambda_{\rm s} \Bigl[\tilde{K}(p_{\rm
s}) + {\cal V} \Bigr] \sin(p_{\rm s} x)}{1+\xi_\Si^2 p_{\rm s}^2 +
\xi_\ka^4 p_{\rm s}^4}
\uqe
with $\tilde{K}(p_{\rm s}) > 0$ from (\ref{Klimits}), while $U$ obeys
\eqan{adsiny2}
\frac{U}{\Si} &=& \frac{U_0}{\Si} - \frac{ \Lambda_{\rm s}^2}{4\xi_\Si^2} 
 \left[ 1 - \frac{|\tilde{K}(p_{\rm s})+{\cal V}|^2}
 {1+p_{\rm s}^2 \xi_\Si^2+p_{\rm s}^4 \xi_\ka^4} \right] 
\naqe

This latter result is plotted in Fig.\ \ref{fig3}(b) and deviates
substantially from the Deryagin prediction as $p_{\rm s}$ increases.
Non-local effects are important and can strongly decrease the membrane
potential energy. In particular, notice that for small $p_{\rm s}$ the
adhesion energy is {\it increased} above the value taken for the planar
situation. This can never occur in the Deryagin approximation (see
(\ref{deryad}) and Ref.\ \cite{us}). The additional Van der Waals
contribution accounted for substantially raises the membrane potential
energy. For large $p_{\rm s}$, positive non-local effects ``saturate''
i.e.\ $\tilde{K}(p_{\rm s})$ plateaus, and $U$ starts to decrease for
greater $p$ values, see (\ref{adsiny2}). In general, the membrane finds it
more difficult to adhere to the rough surface and the adhesion energy will
asymptotically (large $p_s$) tend to a constant value less than unity
(though for the high value of $\Lambda_{\rm s}$ chosen our perturbation
theory is not sufficiently accurate to capture this). The membrane
potential energy eventually becomes positive while its bending energy
vanishes as $p_s \goto \infty$.

Again, it is important to reiterate that the our approach is strictly only
valid when (\ref{perturbvalid}) holds.

\subsection{Comparison between chemically structured and rough surfaces}
\label{comparisonchem}

Consider once more (\ref{elsoln}), with $Q= \xi_\Si^2 q^2 + \xi_\ka^4 q^4$,
\equ
\tilde{\delta h}(\q) = \left\{ \begin{array}{ll} \frac{\Bigl[
\tilde{K}(\q) + {\cal V} \Bigr] \tilde{\zs}(\q)}{1+Q} &
\mbox{\hspace*{8mm} rough} \\ \frac{\tilde{G}(\q) \tilde{\phi}(\q)}{1+Q} &
\mbox{\hspace*{8mm} chemical} \end{array} \right.
\uqe

Two points are worth making; (i) from (\ref{Glimits}) and (\ref{Klimits}).
It can be seen that the membrane amplitude is increased and
 decreased by non-local effects for roughness and chemical structure,
 respectively. This is reflected in a corresponding change
 in the bending energy (compared to the Deryagin results).
 (ii) That due to the different signs of $\tilde{K}+{\cal V}$
 and $\tilde{G}$, see (\ref{Glimits}) and (\ref{Klimits})
 again, the membrane profile is always in phase with the
 surface contour of a rough substrate but is exactly out of phase with
surface structure arising
 due to chemical variation. A membrane adhering
 to a rough surface will, in order to maximize its
 potential energy, try and follow that surface as
 best it can (limited only by the resulting bending
 energy cost). Similarly, the membrane will follow
 the hills and valleys of a substrate potential
 generated by chemical patterning. In this case,
 however, a local increase/decrease in the Hamaker
 constant, $A$, makes that region of the surface
 more/less attractive and so shift the membrane
 in/out. Note that the membrane would be in phase
 with the substrate for a repulsive Van der Waals
 potential \cite{disjoin}.

By using the Deryagin approximation, simple arguments describing an
effective potential energy highlight the different adhesive properties of
the two surfaces. In Fig.\ \ref{fig4}, a surface (translationally
invariant in the $y$-direction) patterned with a square wave profile and a
similar on/off chemical patchwork is schematically shown. For the former
case, a definition for an effective potential energy is given by
\equ
V_{\rm eff} = \frac{\int dx \, V(h-\zs)}{\int dx}
\uqe
which just equals $V(h_0)$ for the planar scenario, $\zs = 0$.
When the substrate is square-corrugated this approximately becomes
\eqan{rfluc}
V_{\rm eff} & \approx & \half \left\{ V(h_0+\delta h - c) + V(h_0- \delta
h + c) \right\} \nonu \\
& = & V(h_0) + \half (\dH-c)^2 V''(h_0) +O \Bigl( (\dH - c)^4 \Bigr) \nonu \\
& > & V(h_0)
\naqe
as $V''(h_0)$ is positive ($h_0$ being a minimum of $V$). Thus
even in this crude argument, one can see that surface roughness
acts (at least if the roughness does not get too large when our
perturbation expansion breaks down) to increase the membrane
potential energy --- a result verified by the Deryagin version of
(\ref{PErough}). Equation (\ref{rfluc}) shows that surface
roughness is reminiscent of Gaussian thermal fluctuations and
similarly to these acts to drive the membrane out of its potential
minimum.

When the wall is patterned with a periodic array of
alternating chemical patches as in Fig.\ \ref{fig4},
one can use similar arguments to those given above to
estimate the effective potential energy. Let the
Hamaker constant $A$ obey
\equ A= \left\{ \begin{array}{l} A_0 (1+\phi) \mbox{\hspace*{10mm}
dark patch} \\ A_0 (1-\phi) \mbox{\hspace*{10mm} light patch}
\end{array} \right.
\uqe
for constant and positive $A_0$ and $\phi$, then the potential
energy is, see (\ref{Afull}) and (\ref{nochein}),
\equ
V(h) = V_{\rm vdw}(h) \cdot \as(x) + V_{\rm hyd}(h)
\uqe
with $\as(x) = 1 \pm \phi$ on the dark/light patches. Therefore
\eqan{Vpotchem}
V_{\rm eff} &=& \frac{\int dx \, V(h)}{\int dx} \nonu \\
&=& \half \Bigl\{ V_{\rm vdw}(h_0-\dH) \cdot (1+\phi) + V_{\rm
hyd}(h_0-\dH) \nonu \\
& & + V_{\rm vdw}(h_0+\dH) \cdot (1-\phi) + V_{\rm hyd}(h_0+\dH) \Bigr\}
\nonu \\
&=& V(h_0) + \half \dH^2 V''(h_0) - \phi \dH V_{\rm vdw}'(h_0) + O( \dH^3 )
\naqe
Notice here that the membrane's position is exactly out
of phase with the surface structure. This leads to a negative
contribution in (\ref{Vpotchem}) and one can see that the new
membrane configuration, as the negative term is of order $\dH$ and
positive is of order $\dH^2$, is likely to result in a net attractive
contribution to the potential energy. Indeed, this can be verified
by summing (\ref{chemPE}) and (\ref{chemBE}) which, as already
mentioned, is always negative.

In summary, a membrane generally adhers less favorably (relative to the
planar and homogeneous surface) to a rough substrate and adopts a
configuration that is in phase with the surface contours. A chemically
structured substrate has a higher adhesion energy (more favorable) and
leads to a membrane configuration exactly out of phase with the surface
structure.

\section{Comparison between exact and approximate solutions}
\label{results2}

From these examples and those given in Ref.\ \cite{us}, it is clear that
the Deryagin approximation is certainly the most versatile if one wishes
to obtain analytical results. However, it is also apparent that in some
 situations non-local effects can become important and in this section
 we compare the Deryagin result with an exact numerical solution.

The scenario we choose to specialize to involves solely geometric
inhomogeneities. We consider a chemically pure substrate made up of a
regular array (in the $x$--direction) of `V' shaped trenches. These could
be formed, for example, by etching silicon wafers \cite{gerdes}. In the
other spatial dimension, i.e.\ the $y$--direction, the system is
translationally invariant. For simplicity, the trenches are assumed to be
symmetric about their lowest point and have a maximum width of $d$ and a
depth of $\lambda d$.  See Fig.\ \ref{fig5} for an example.

\subsection{The Deryagin solution}

To find the Deryagin solution it is most easy to consider
(\ref{el2}) directly, which reduces to a one dimensional differential
equation. This is
\equn{el45}
\left( \xi_\kappa^4 \frac{d^4}{dx^4}-\xi_\Si^2 \frac{d^2}{dx^2} +
1 \right) \delta h(x) = \zs(x)
\nuqe
with
\eqan{zsig}
\zs(x) &=& 2 \lambda \sum_{n=-\infty}^{\infty} \Biggl\{ r_x
\Bigl[ \theta(x+nd)-\theta(x+nd-d/2) \Bigr] \nonu \\
& & + (d-r_x) \Bigl[ \theta(x+nd -d/2) - \theta(x+n d-d) \Bigr] \Biggr\}
\naqe
where $\theta(x)$ is the Heaviside function and $r_x = x \, {\rm mod} \,
d$, i.e.\ the remainder of $x$ on division by $d$. Assuming that the
membrane has the same symmetry as the substrate, the boundary conditions
are
\eqan{conditions}
h'(n d) &=& h'''(n d) =0 \nonu \\
h'(n d+x)&=& h'''(n d +x)=0
\naqe
for all integer $n$.

Due to the periodicity of $\zs(x)$, we need only solve
(\ref{el45}) for $0<x<d/2$, where $\zs(x)=2\lambda x$, and then
reflect and/or translate this solution to obtain the full membrane
configuration. We find in this region that for $\y = (x,y)$,
\equn{el45soln}
h(\y) = h_0 + 2\lambda x - \frac{2 \lambda \eta_+^2 \eta_-^2}
{\eta^2_+ - \eta_-^2} \left[ \psi_+(x) - \psi_-(x) \right]
\nuqe
with
\equ
\psi_\pm(x) = \frac{{\rm sinh} \Bigl[ (d/4-x)\eta_\pm \Bigr]}
{\eta_\pm^3 {\rm cosh} \Bigl[(\eta_\pm d/4 \Bigr)}
\uqe
The $\eta_\pm$ come from the factorization of the operator in
(\ref{el45}), see Ref.\ \cite{us}, and are
\equn{lams}
\eta_\pm = \xi^{-1} \left[ \frac{1 \pm \sqrt{1-4\chip^4}}{2}
\right]^\half
\nuqe
where $\xi$ is given by (\ref{xidefn}). The height profile of the
membrane, $h(x)$ is plotted  in Fig.\ \ref{fig5} using
(\ref{el45soln}).

The adhesion energy is obtained from the definition (\ref{ad}).
Equation (\ref{f2}) with (\ref{spikes}) implies
\equ
\frac{U}{\Si} = \frac{U_0}{\Si} - \frac{1}{d} \int_0^{d/2} dx
\Bigl\{ {h'}^2 + \xi^2 {h''}^2 + \xi_\Si^{-2} (h - h_0 - 2 \lambda x)^2 \Bigr\}
\uqe
Using (\ref{el45soln}), the above integral can be calculated analytically
yielding
\equn{uzigzag}
\frac{U}{\Si} = \frac{U_0}{\Si} - 4 \lambda^2 \left\{ \half +
\frac{I(d \eta_+, d \eta_-) + I(d \eta_-, d \eta_+)}
{d^4 (\eta_+^2-\eta_-^2)^2} \right\}
\nuqe
with
\eqa
I(u,v) &=& \frac{2u\chip^4}{(\xi/d)^2} \Biggl\{ \frac{4}
{1+\e^\frac{u}{2}} - \frac{2 \Bigl[1+2\chip^2
\Bigr]}{(\xi/d)^2(u^2-v^2)}\nonu \\
& &  + \frac{(\xi/d)^2 v^2(2u^2-v^2)}{\chip^4u^2} {\rm tanh} (u/4) \Biggr\}
\aqe
Fig.\ \ref{fig6} (dashed lines)
illustrates (\ref{uzigzag}) and compares it with the exact
numerical solution detailed below.

\subsection{The exact numerical solution}

In this subsection we present a numerical solution which accounts
for the full Van der Waals interaction (\ref{pot0}) and the
bending energy term in (\ref{f}). One can functionally minimize
(\ref{f}) but the resulting Euler-Lagrange equation is non-linear and very
complicated. Given that (\ref{zsig}) implies the
boundary conditions (\ref{conditions}), the resulting problem is
extremely awkward to tackle even numerically if one tries to solve
the equation directly. Instead, we choose first to discretize
(\ref{f}) and then minimize the free energy functional with
respect to all of the discrete variables. This has the advantage
that the boundary conditions, (\ref{conditions}), can be easily
incorporated.

For one dimensional structure, that is relevant to the
trench geometry, (\ref{f}) becomes
\equn{FfullBE}
{\cal F}[h] = \int dx \Biggl \{ \Si \sqrt{1+{h'}^2} +
\frac{\ka}{2} \cdot \frac{{h''}^2}{(1+{h'}^2)^\frac{5}{2}} +
V(h;\zs,0) \Biggr \}
\nuqe
Due to the symmetry of (\ref{zsig}), we only need consider a
solution for $0<x<d/2$ and adopt a standard discretization
process by dividing the interval, i.e.\
\equ
\begin{array}{lcl}
h'(x_k) \leftrightarrow \frac{h_{k+1}-h_k}{\Delta} &\hspace*{5mm} ;
\hspace*{5 mm} & h''(x_k)
\leftrightarrow \frac{h_{k+2}-2h_{k+1}+h_k}{\Delta^2}
\end{array}
\uqe
where $x_k = k \Delta$, $k=1, \dots,N$ and $\Delta = \frac{d}{2N}$.
Here, $N$ is the number of points making up the
one-dimensional lattice and typically was chosen
somewhere between $100$ and $400$. For the simple
surface (\ref{zsig}), (\ref{pot}) can be broken
into a (convergent) infinite sum of integrals each
of which can then be evaluated analytically.

Carrying out this procedure we find Figs. \ref{fig6} and \ref{fig7}. From
the adhesion energies plotted in Fig.\ \ref{fig6}, there is a region of
good agreement between the Deryagin and exact solution. This agreement
occurs for low $d$ and $\lambda$ where (\ref{perturbvalid}) holds. As the
Deryagin approximation underestimates the attractive potential in this
case, the adhesion energy is always less than the exact result and it is
also reassuring that the exact $U$ does not vanish for high $d$ or
$\lambda$. This prediction of the Deryagin approach is clearly an artifact
of going beyond the limits of perturbation theory. Generally, we see that
the adhesion energy decreases with greater values of the roughness, i.e.\
large $\lambda$ or $d$ (for small $d$ any structure is effectively `washed
out' and not seen by the membrane).

The degree of penetration of the membrane into the `V'-shaped trenches is
shown in Fig.\ \ref{fig7} where the membrane height above the middle of
the trench, $\dH(0)= h(0)-h_0$, is plotted. One can see that the membrane
always lies further away from the substrate than if the latter were
entirely planar and so there is no penetration into the surface
indentations. However, this could be encouraged by having flat regions
separating each trench (see Ref.\ \cite{us} for a similar example). As the
surface becomes rougher, it also becomes more repulsive ($U$ decreases)
and so the membrane moves outwards. It is perhaps surprising to see that
the exact solution lies furthest away from the substrate despite having a
higher adhesion energy. This is likely to be a consequence of higher order
bending energy terms in (\ref{FfullBE}) reducing the amplitude of the
membrane configuration and so increasing its height at the center of a
trench.

\section{Conclusion}

 A significant experimental question is whether or not
substrate structure encourages membrane adhesion. An important
conclusion of our study is that chemical structure always
increases a substrate's attractiveness; the membrane potential
energy (\ref{chemPE}) is clearly negative and also greater than
the bending energy, (\ref{chemBE}). Consequently, the adhesion
energy increases. Rough surfaces are unfortunately more ambiguous 
as (\ref{PErough}) is not of a definite sign. However, if the
Deryagin approximation is invoked ($\tilde{K}(q)+{\cal V} \approx
1$), then the potential energy contribution is always positive.
Therefore, we would expect roughness to usually decrease a
substrate's attractiveness and lead to a drop in the adhesion
energy. We should emphasize that for surfaces for which non-local
effects are important this may no longer be the case.

Finally, some comments are pertinent on the validity of our approach. Both
analytical methods breakdown when the amplitude of the structure, be it
geometrical or chemical, becomes large. This is to be expected as our
analysis is fundamentally a perturbation method and can only be
confidently followed when (\ref{perturbvalid}) holds. The linear response
technique is an improvement over Deryagin and is particularly appropriate
for rough surfaces where the additional non-local effects lead to an
increase in the amplitude of an adhering membrane. For smoothly varying
surfaces these effects can even lead to the surface becoming attractive
--- a result that is not predicted by the local Deryagin approximation.
Unfortunately, it is difficult to identify the particular geometries for
which non-local Van der Waals contributions are important but for those
surfaces of biotechnological interest, i.e.\ with trenches or indentations
etched into them, they do not seem to lead to radically different
behavior. To conclude, if one wishes an analytical guide to how a certain
substrate structure will influence membrane adhesion and if that structure
can be conveniently described in Fourier space then the linear response
description is the method of choice. Failing this the Deryagin
approximation is quick and easy to apply if only normally adequate for
small amplitude effects.


\acknowledgments

We are particularly grateful to J.\ R\"adler and E.\ Sackmann for
introducing us to the problem of membrane adhesion on rough surfaces, for
numerous discussions and suggestions, and for sharing with us their
experimental results.  We benefited from conversations and correspondence
with P.\ Lenz, U.\ Seifert and P.\ B.\ Sunil Kumar.  Partial support from
the Israel Science Foundation founded by the Israel Academy of Sciences
and Humanities --- centers of Excellence Program and US--Israel Binational
Science Foundation (BSF) under grant no.\ 98-00429 is gratefully
acknowledged.



\begin{table}[p]
\begin{tabular}{llll}
$\ka = 35T$ & $\Si = 1.7 \times 10^{-5} \; {\rm Jm}^{-2}$
& $\delta = 38\; \mbox{\AA}$ &  $T = 4.1 \times 10^{-21}\; {\rm J}$\\
${A_0} = 2.6 \times 10^{-21} \;
{\rm J}$ & $b =0.93 \; {\rm Jm}^{-2} $ & $\alpha^{-1} =2.2 \; \mbox{\AA} $& \\
$a \simeq 49.3 \; \mbox{\AA}$ &
$h_0 \simeq 0.61 a\simeq 30\; \mbox{\AA}$
& $U_0 \simeq 0.298 \Si$ & $v \simeq 22.85 a^{-2} \Si $ \\
 $ \xi_\Si \simeq 0.21a $ & $\xi_\ka \simeq 1.97 a
 $ & $\xi \simeq 18.62a$ & ${\cal V} \approx 1.30$
\\
\end{tabular}
\caption{The various parameters, chosen and calculated,
for a supported membrane. For definitions see text.}
\label{tab}
\end{table}


\newpage

\section*{Figure Captions}

\begin{figure}[tbh]
\vspace{3cm} \centerline{ \epsfxsize=16cm
\hbox{\epsffile{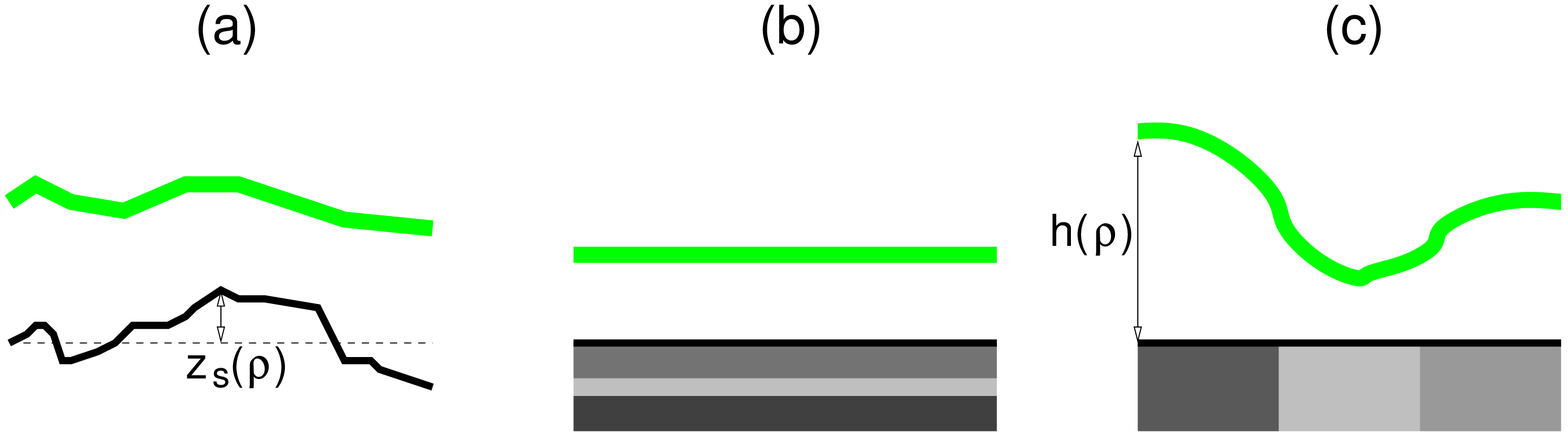}}} \vspace{8cm}
\caption[]{}%
\label{fig1}
\end{figure}

\begin{figure}[tbh]
\vspace{3cm} \centerline{ \epsfysize=10cm
\hbox{\epsffile{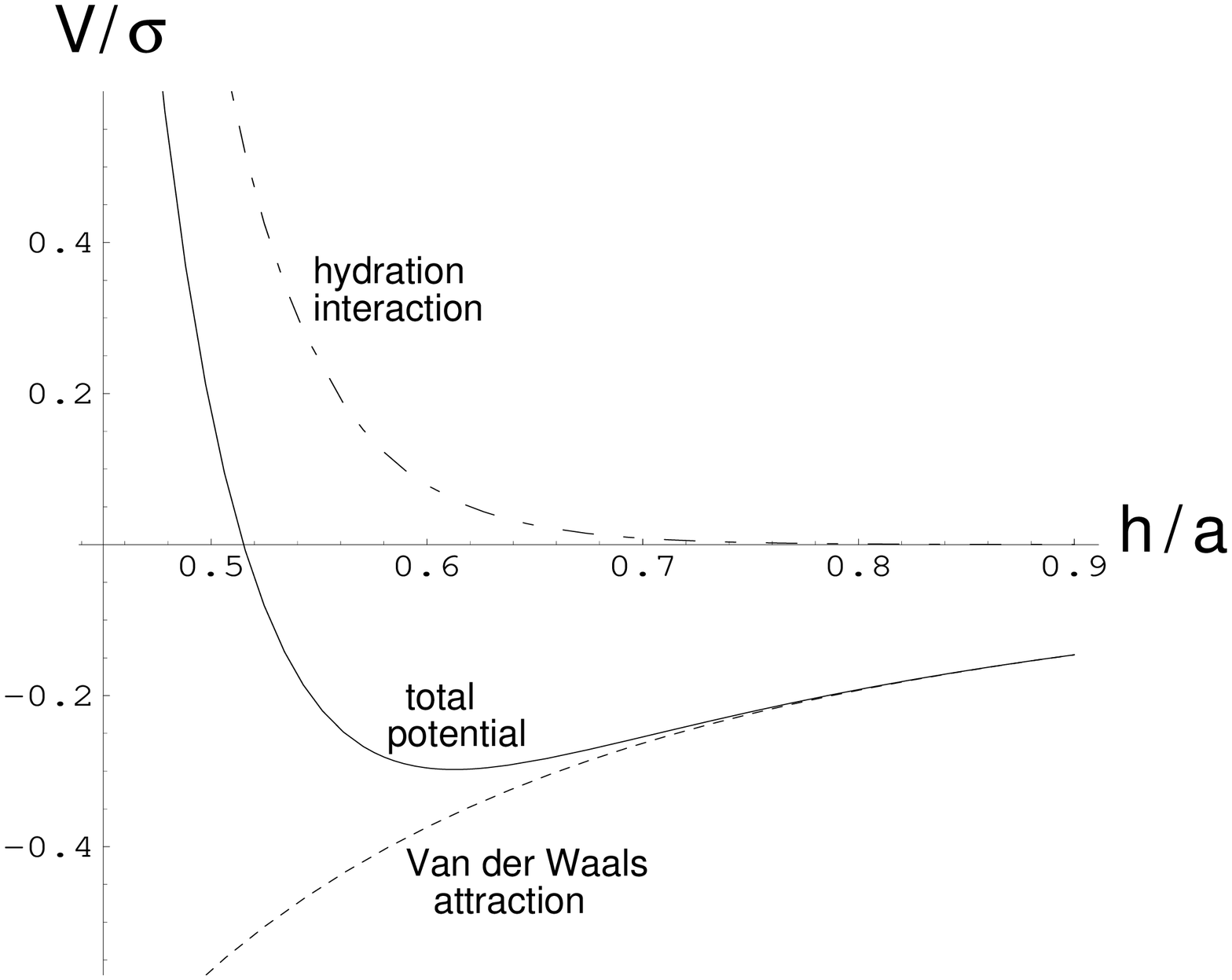}}} \vspace{3cm}
\caption[]{}%
\label{fig2}
\end{figure}

\begin{figure}[tbh]
\vspace{3cm} \centerline{ \epsfysize=8cm
\hbox{\epsffile{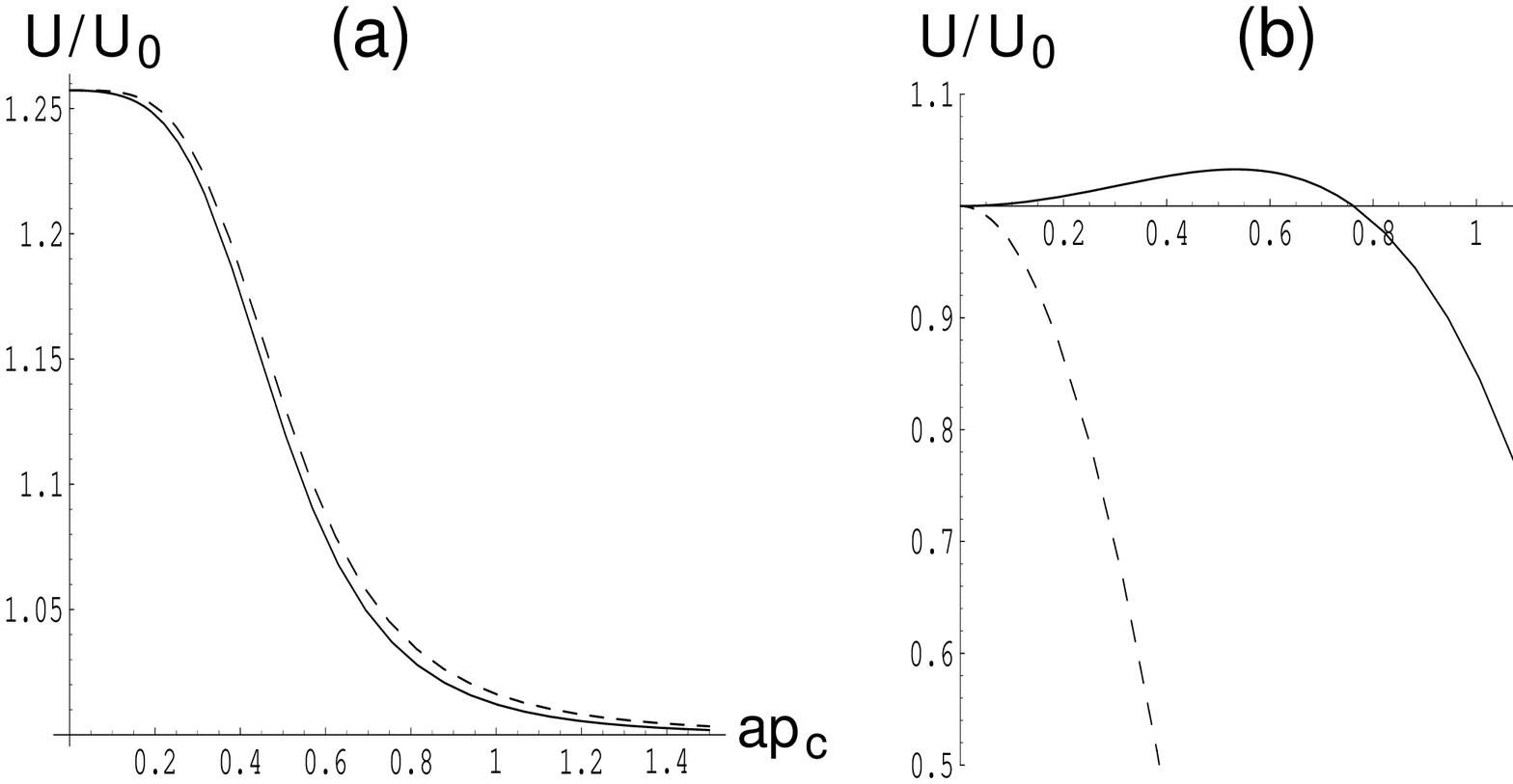}}} \vspace{3cm}
\caption[]{}%
\label{fig3}
\end{figure}

\begin{figure}[tbh]
\vspace{3cm} \centerline{ \epsfysize=10cm 
\hbox{\epsffile{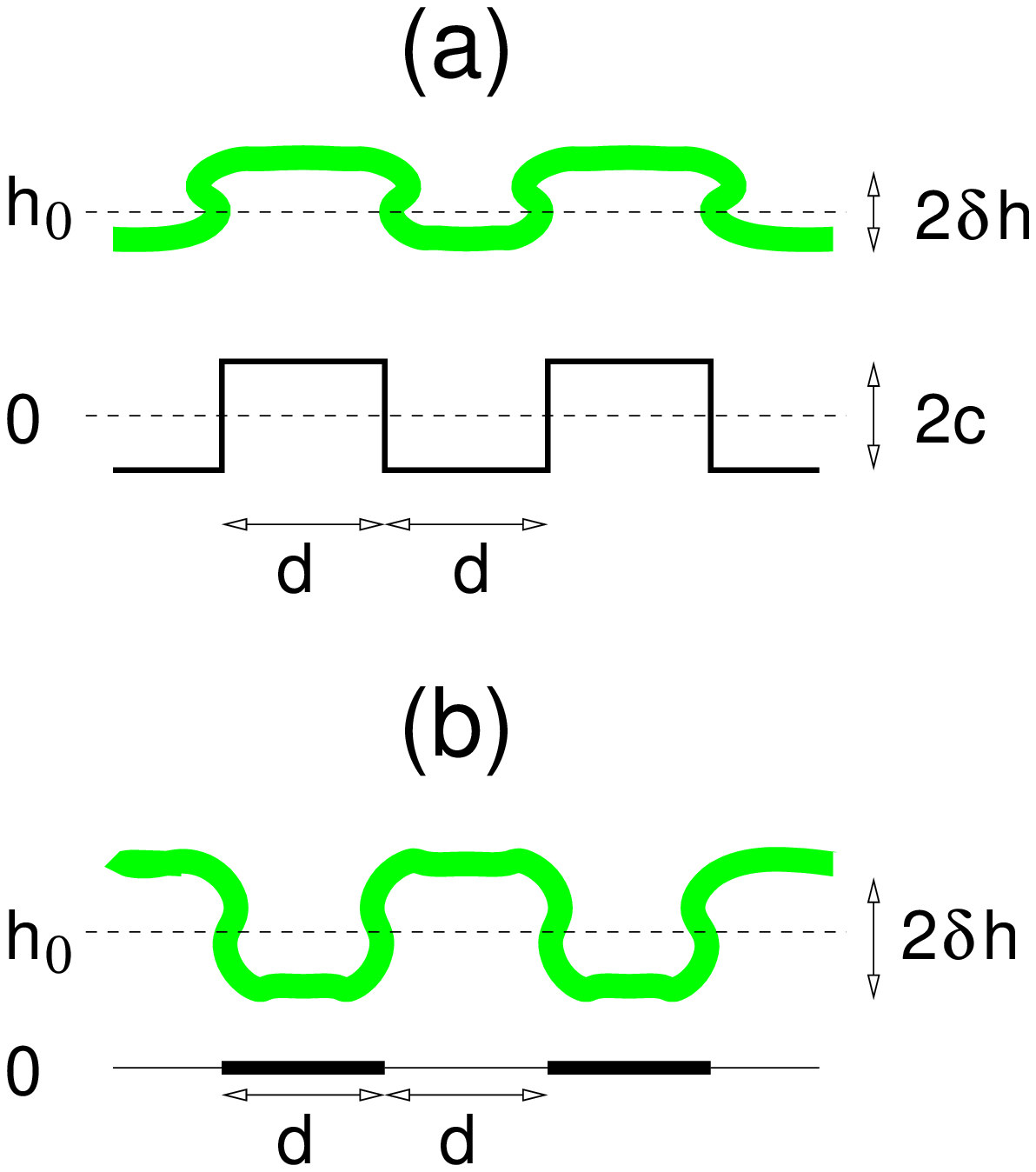}}} \vspace{3cm}
\caption[]{}%
\label{fig4}
\end{figure}

\begin{figure}[tbh]
\vspace{3cm} \centerline{\epsfysize=8cm 
\hbox{\epsffile{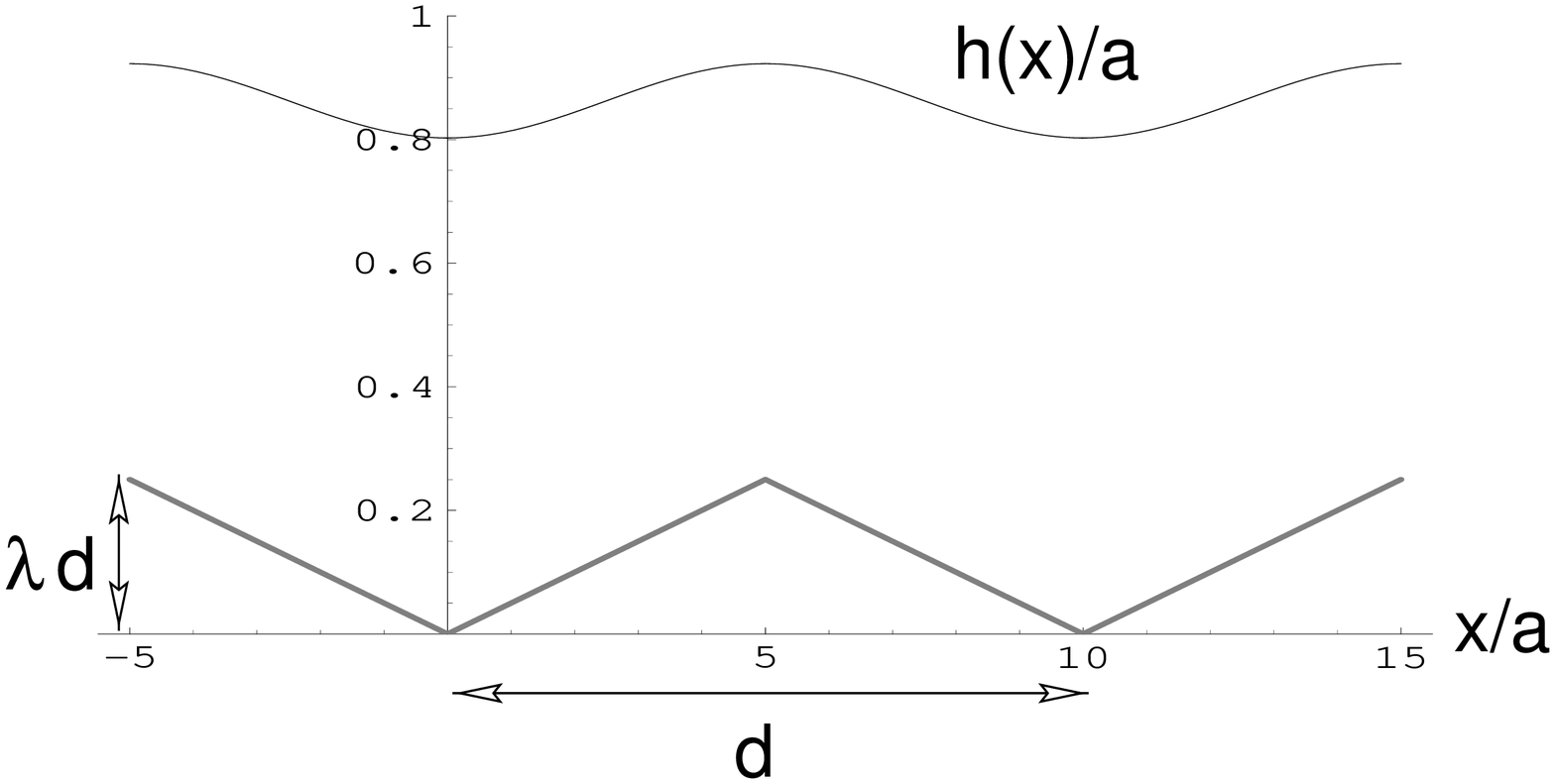}}} \vspace{3cm}
\caption[]{}%
\label{fig5}
\end{figure}

\begin{figure}[tbh]
\vspace{3cm} \centerline{\epsfysize=12cm \epsfxsize=18cm
\hbox{\epsffile{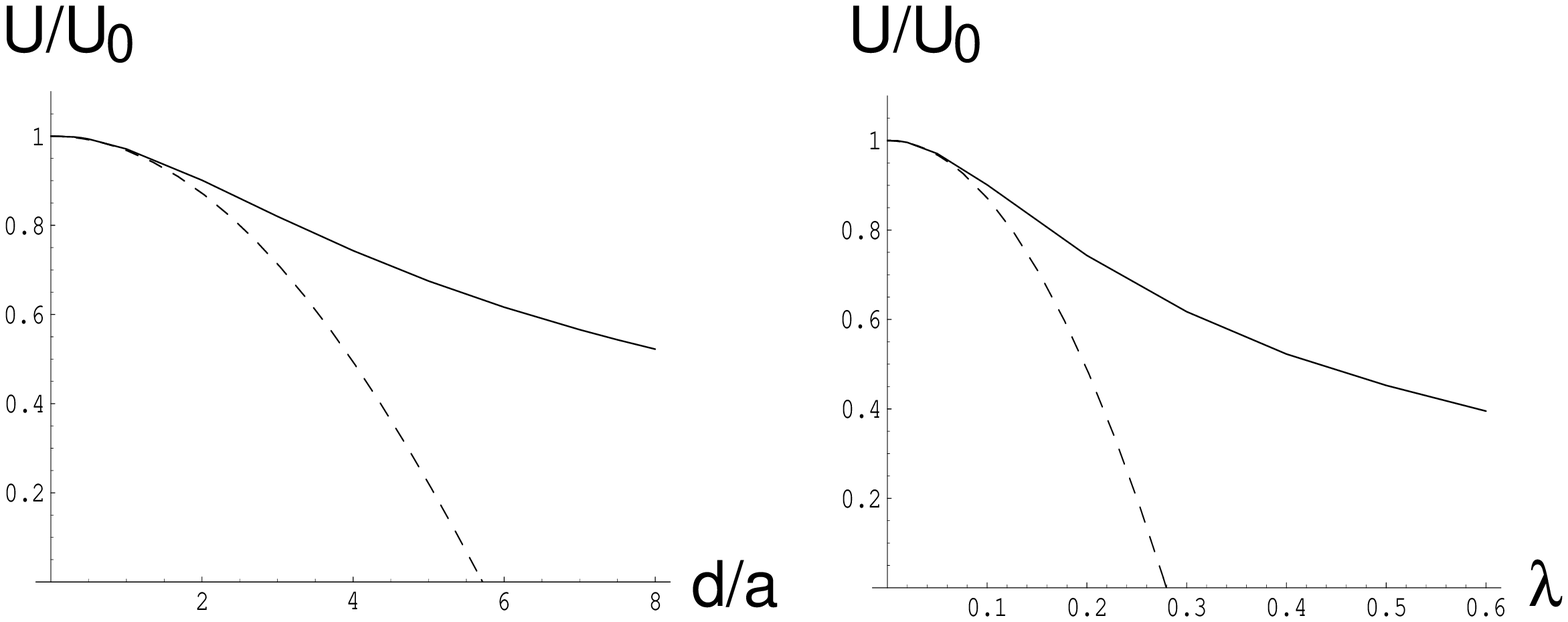}}} \vspace{3cm}
\caption[]{}%
\label{fig6}
\end{figure}

\begin{figure}[tbh]
\vspace{3cm} \centerline{ \epsfysize=9cm \epsfxsize=16cm
\hbox{\epsffile{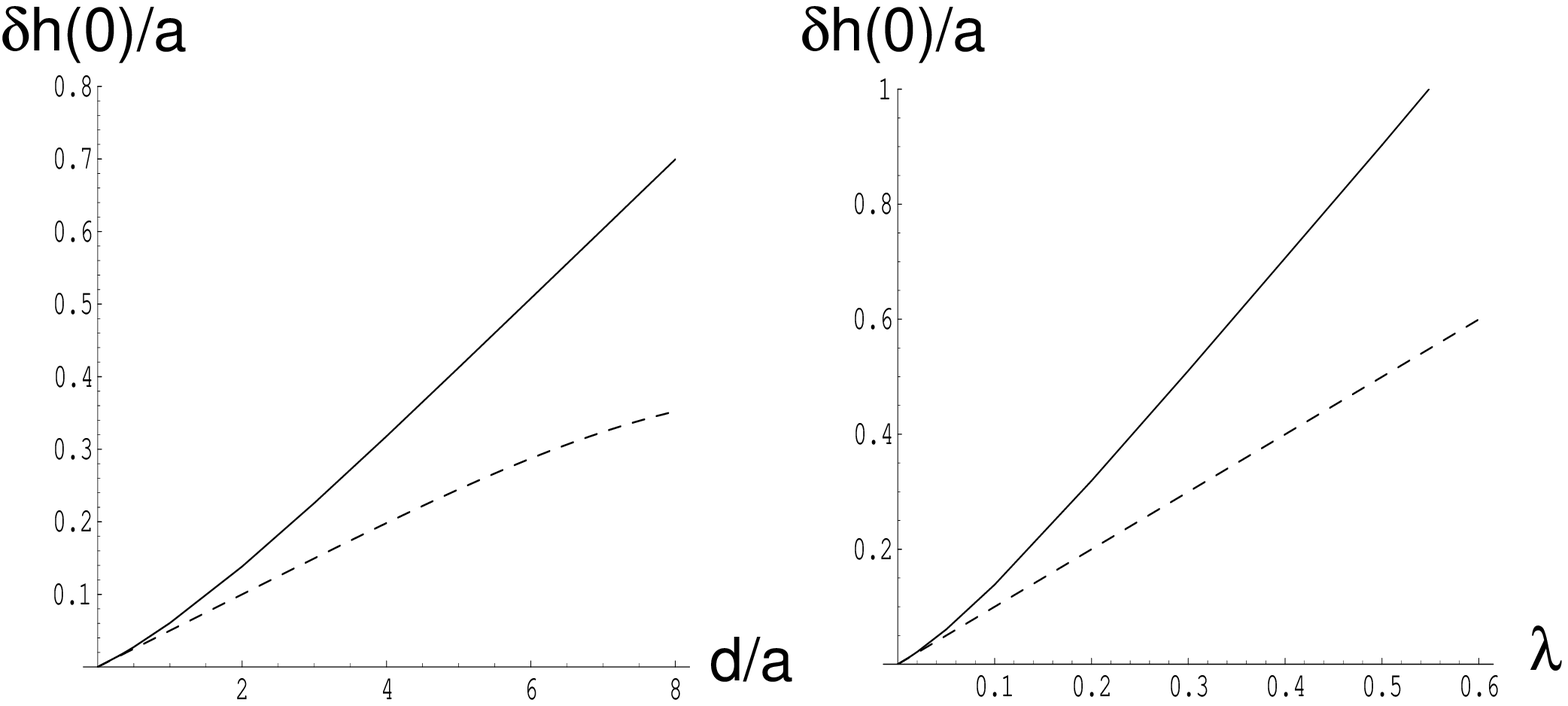}}} \vspace{3cm}
\caption[]{}%
\label{fig7}
\end{figure}

\begin{itemize}

\item[{\bf Figure~\ref{fig1}}]
Supported membranes on structured surfaces. In (a), the membrane is
adhering to a rough but otherwise homogeneous surface. The reference
$\y$-plane is shown as a dashed line. The height of the lower membrane
lipid leaflet and the surface, measured from this plane, are denoted by
$h(\y)$ and $z_\rw(\y)$, respectively. Layered and columnar chemical
structure are sketched in (b) and (c), respectively.

\item[{\bf Figure~\ref{fig2}}]
A plot of the various interactions described in Sec.\ \ref{planarand},
with parameter values given in Table \ref{tab}. Here all potentials are
measured in units of the membrane tension, $\sigma = 1.7 \times 10^{-5}
{\rm Jm}^{ -2}$, and lengths in terms of $a \simeq 49.3$\AA. The total
potential has a minimum at $h_0 \simeq 0.6a$.

\item[{\bf Figure~\ref{fig3}}]
Membrane adhesion energy for a substrate which (a) is flat and has a
sinusoidally varying Hamaker constant ($\Lambda_{\rm c}=10.0$) and (b) has
a sinusoidal surface configuration ($\Lambda_{\rm s}=2.0 a$) plotted
against wavenumber. The Deryagin solution is also shown with a dashed
line.

\item[{\bf Figure~\ref{fig4}}]
A membrane adhering to (a) a substrate with a square wave surface
configuration and (b) to a flat substrate periodically striped with two
different chemical compounds.

\item[{\bf Figure~\ref{fig5}}]
A typical membrane configuration predicted by the Deryagin approximation
for adsorption above a homogeneous substrate sculptured with `V'-shaped
trenches. The parameters for a single trench are $d=10a$ and $\lambda =
0.05$, which implies a width of around $500\,$\AA \ and depth of
approximately $25\,$\AA. The adhesion energy is $U \approx 0.43 U_0$.

\item[{\bf Figure~\ref{fig6}}]
A comparison of the Deryagin predictions (dashed lines) and an exact
numerical solution (heavy lines) for the adhesion energy, $U$, above a
substrate patterned with `V' shaped trenches. In (a) $U/U_0$ is plotted as
a function of $d/a$ with $\lambda = 0.1$, while for (b), it is shown as a
function of $\lambda$ and $d=2a$. For small $d/a$ or $\lambda$ the
Deryagin method provides a good approximation to the numerical result.

\item[{\bf Figure~\ref{fig7}}]
The membrane height in the center of a `V' shaped trench measured with
respect to its planar height, $\delta h(0)=h(0)-h_0$. In (a) $\dH(0)$ is
plotted as a function of the trench width $d/a$ while in (b) it is plotted
as a function of the trench amplitude with $d=2a$. The numerical solution
(heavy line) soon departs from the Deryagin result (dashed line) with good
agreement again only occurring for values of the adhesion energy close to
$U_0$. %
 
\end{itemize}

\pagebreak

{\hfill P. Swain \& D. Andelman}
\newpage

{\hfill P. Swain \& D. Andelman}
\newpage

{\hfill P. Swain \& D. Andelman}
\newpage

{\hfill P. Swain \& D. Andelman}
\newpage

{\hfill P. Swain \& D. Andelman}
\newpage

{\hfill P. Swain \& D. Andelman}
\newpage

{\hfill P. Swain \& D. Andelman}
\newpage


\begin{references}
\bibitem{gb} The Gauss-Bonnet theorem then implies that the Gaussian
curvature contribution is constant and so can be safely neglected.
\bibitem{disjoin} Note that this is the usual wetting case where a
negative Hamaker constant leads to a repulsive Van der Waals force -- the
disjoining pressure of Deryagin. In fact, to recover a wetting scenario in
our equations one needs simply to take the limit of the membrane thickness
$\delta$ going to infinity.
\end{references}
\end{document}